\documentclass[a4paper,UKenglish,cleveref, autoref, thm-restate]{lipics-v2021}
\usepackage{nicematrix}
\usepackage[T1]{fontenc}
\usepackage{amsmath}
\usepackage{amssymb}
\usepackage{fancyvrb}       
\usepackage[most]{tcolorbox}
\usepackage{xcolor}
\usepackage{tikz}
\usepackage{enumitem}
\usepackage{hyperref}
\usepackage{graphicx}
\usepackage{everysel}
\usepackage{float}
\usepackage{caption}
\usepackage{gensymb}
\usepackage{tikz}
\usepackage{subfiles}
\usepackage[normalem]{ulem}
\usetikzlibrary{arrows.meta}
\usetikzlibrary{matrix,fit}
\usetikzlibrary{patterns}
\tikzset{ 
table/.style={
  matrix of nodes,
  row sep=-\pgflinewidth,
  column sep=-\pgflinewidth,
  nodes={rectangle,draw=black,text width=1.25ex,align=center},
  text depth=0.25ex,
  text height=1ex,
  nodes in empty cells
  },
texto/.style={font=\footnotesize\sffamily},
title/.style={font=\small\sffamily}
}

\newcommand{\FOA}{\mathsf{FOA}}
\newcommand{\FOB}{\mathsf{FOB}}
\newcommand{\Fair}{\mathsf{Fair}}
\newcommand{\UM}{\mathsf{UM}}
\newcommand{\MM}{\mathsf{MM}}

\title{Verified Double Sided Auctions for Financial Markets} 

\author{Raja Natarajan}{Tata Institute of Fundamental Research, Mumbai, India.} {raja@tifr.res.in}{}{}
\author{Suneel Sarswat}{Tata Institute of Fundamental Research, Mumbai, India.} {suneel.sarswat@gmail.com}{}{}
\author{Abhishek Kr Singh}{Birla Institute of Technology and Science Pilani, Goa, India.} {abhishek.uor@gmail.com}{}{}

\authorrunning{N. Raja, S. Sarswat, and A. Singh}

\Copyright{Raja Natarajan, Suneel Sarswat, and Abhishek Kr. Singh}

\ccsdesc[500]{Information systems~Online auctions}
\ccsdesc[500]{Software and its engineering~Formal software verification}
\ccsdesc[300]{Theory of computation~Algorithmic mechanism design}
\ccsdesc[300]{Theory of computation~Computational pricing and auctions}
\ccsdesc[100]{Theory of computation~Program verification}
\ccsdesc[100]{Theory of computation~Automated reasoning}

\keywords{Double Sided Auction, Formal Verification, Financial Markets, Proof Assistant}

\allowdisplaybreaks[1]

\begin{document}

\nolinenumbers
\maketitle

\begin{abstract}

Double sided auctions are widely used in financial markets to match demand 
and supply. Prior works on double sided auctions have focused primarily on single quantity
trade requests. We extend various notions of double sided auctions to incorporate multiple
quantity trade requests and provide fully formalized matching algorithms 
for double sided auctions with their correctness proofs. We establish new uniqueness 
theorems that enable automatic detection of violations in an exchange program by comparing 
its output with that of a verified program.
All proofs are formalized in the Coq proof assistant without adding any
axiom to the system. We extract verified OCaml and Haskell programs that can be used 
by the exchanges and the regulators of the financial markets.
We demonstrate the practical applicability of our work by running the 
verified program on real market data from an exchange to automatically 
check for violations in the exchange algorithm.

\end{abstract}

\section{Introduction}
\label{sec:introduction}

Computer algorithms are routinely deployed nowadays by all big stock exchanges 
to match buy and sell requests. These algorithms are required to abide by 
various regulatory guidelines. 
For example, market regulators make it mandatory for trades resulting from
double sided auctions at exchanges to be fair, uniform and individual-rational.

In this paper, we introduce a formal framework for analyzing trades 
resulting from double sided auctions used in the financial 
markets. To verify the essential properties required by market regulators, 
we formally define these notions in a theorem prover and then 
develop important results about matching demand and supply. Finally, we 
use this framework to verify properties of two important classes of double 
sided auction mechanisms. 

One of the resulting advantages of our work for an exchange or a regulator
is that they can check the algorithms deployed for any violations from required
properties automatically. This is enabled by the new uniqueness results that we establish in this work.
 All the definitions and results presented in this 
paper are completely formalized in the Coq proof assistant without adding 
any additional axioms to it.
The complete formalization in Coq facilitates automatic program 
extraction in OCaml and Haskell, with the guarantee that extracted programs satisfy
the requirements specified by the market regulator. 
Consequently, the extracted program could also be deployed directly at 
an exchange, apart from checking for violations in existing programs.
We demonstrate the practical applicability of our work by running the 
verified program on real market data from an exchange to automatically 
check for violations in the exchange algorithm.

The rest of this paper is organized as follows: 
Section \ref{sec:background} provides a brief background and overview of trading at exchanges which is needed to
describe our contributions;
In Section \ref{sec:contribution}, we briefly state our contributions;
Section \ref{sec:modeling} provides basic definitions and establishes certain combinatorial results;
Section \ref{sec:fair} describes a fairness procedure;
Section \ref{sec:uniform} describes the uniform matching mechanism used in the financial markets; 
Section \ref{sec:maximum} describes the maximum matching mechanism; 
Section \ref{sec:uniqueness} establishes uniqueness results that enables automatic checking for violations in an exchange matching algorithm;
Section \ref{sec:demonstration} describes the practical utility of our work through running our verified program on real market data from an exchange;
Section \ref{sec:conclusion} concludes the paper with related work and future directions.
Parts of some sections have been moved to the appendix.


\section{Background}
\label{sec:background}
Financial trades occur at various types of exchanges. For example, there are exchanges for stocks, commodities and currencies. 
At any exchange, multiple buyers and sellers participate to trade certain products. Mostly exchanges employ double sided auction
mechanisms to match the buyers and sellers. Some exchanges, apart from using double sided auctions, also use an 
online continuous algorithm for executing trades during certain time intervals, especially for highly traded products. 

For conducting trades of a certain product using a double sided auction mechanism, the exchange collects buy and sell requests 
from the traders for a fixed time period. At the end of this time period, the exchange matches some of the trade requests and 
outputs trades, all at a single price. This price is sometimes referred to as the equilibrium price and the process as price
discovery. A buyer places a buy request, also known as a \emph{bid}, which consists of a quantity indicating the 
maximum number of units he is interested in buying and a common maximum price (bid's limit price) for each of the units. Similarly, 
a seller's sell request, an \emph{ask}, consists of a quantity and a minimum price (ask's limit price). Each trade (\emph{transaction}) consists of
a bid, an ask, traded quantity, and a trade price. Naturally, the traded quantity should be at most the minimum of the bid and the ask quantities and the trade price should be compatible with the bid and the ask. 

Apart from the single price property and compatibility constraint mentioned above, 
there are other desired properties that the trades (\emph{matching}) should have. The properties 
that capture these constraints are: \emph{uniform}, \emph{individual-rational}, \emph{fair} and \emph{maximum}. We briefly describe these matching properties:

\begin{itemize}
\item{\textbf{Uniform}:} A matching is uniform if all the trades happen at the same price.
\item{\textbf{Individual-rational}:} A matching is individual-rational if for each matched bid-ask pair the trade price is between the bid and ask limit prices. In the context of financial markets, the trade price should always be between the limit prices of the matched bid-ask pair.
\item{\textbf{Fair}:} A bid $b_1$ is more competitive than a bid $b_2$ if $b_1$ has a higher limit price than $b_2$ or if their limit prices are the same and $b_1$ arrives earlier than $b_2$. Similarly, we can define competitiveness between two asks. A matching is \emph{unfair} if a less competitive bid gets matched but a more competitive bid is not fully matched. Similarly, it could be unfair if a more competitive ask is not fully matched. If a matching is not unfair, then it is fair. 
\item{\textbf{Maximum}:} A matching is maximum if it has the maximum possible total traded quantity among all possible matchings.
\item{\textbf{Optimal individual-rational-uniform}:} An individual-rational and uniform matching is called optimal individual-rational-uniform if it has the  largest total trade volume among all matchings that are individual-rational and uniform.
\end{itemize}

No single algorithm can possess all the above first four properties simultaneously \cite{WurmanWW98,mcafee1992}. 
In the context of financial markets, regulators insist on the matching being fair and optimal individual-rational-uniform, 
thus compromising on the maximum property. In other contexts where the matching being maximum is important along with individual
rational and fair, uniformity is lost.
This gives rise to two different classes of double sided auction mechanisms, each with a different objective. In our work, we
consider both these classes of mechanisms.

\section{Our Contributions}
\label{sec:contribution}

In this work, we formalize the notion of double sided auctions where trade requests can be of multiple quantities.
Prior to our work, similar notions were explicitly defined only for single unit trade requests \cite{icfem,NiuP13,ZhaoZKP10}.
In going from formalizing the theory for single unit to the theory of multiple units, 
the mechanisms and their correctness proofs changed substantially. Due to the possibility of partial trades, the formal analysis of multiple unit trades becomes significantly more involved than in \cite{icfem}. In this work, we show how to  efficiently handle this extra complexity  by making the functions and their properties sensitive to the partial trade quantities. This helps us to develop formal proofs of correctness of the recursive mechanisms for double sided auctions. 

In addition, we provide new uniqueness results that guarantee that the matching algorithm for the double sided auctions used in the financial markets outputs a unique volume of trades per order if the algorithm is fair and optimal individual-rational-uniform; thus enabling automatic checking of violations in the exchange algorithm by comparing its output with that of a verified program. We demonstrate this by running the extracted OCaml code of our certified mechanism on real data from an exchange and comparing the outputs.
Following is a brief description of the key results formalized in this work.

\begin{itemize}
\item{\textbf{Combinatorial result}:} We show that the modeling and the libraries we created to obtain our results are also useful in proving other important results on double sided auctions. For example, in Theorem \ref{thm:boundM}, we show that for any $p$, no matching can achieve a trade volume higher than the sum of the total demand and the total supply in the market at price $p$.

\item{\textbf{Fairness}:} We show that any matching can be converted into a fair matching without compromising on the total traded volume. For this, we design an algorithm, the $\Fair$ procedure, which takes a matching $M$ as input, and outputs a matching $M'$.
In Theorem \ref{thm:existsFair}, we show that the total traded quantities of $M$ and $M'$ are the same and $M'$ is a fair matching. 

\item{\textbf{Uniform mechanism}:} We design an algorithm, the $\UM$ procedure, that takes as input the bids and the asks and outputs a fair, individual-rational and uniform matching. Furthermore, in Theorem \ref{thm:um}, we show that the output matching 
has the largest total trade volume among all the matchings that are uniform and individual-rational and thus is optimal individual-rational-uniform. This algorithm is used in the exchanges that output trades using double sided auctions.

\item{\textbf{Maximum mechanism}:}  We design an algorithm, the $\MM$ procedure, that takes as input the bids and the asks and outputs an 
individual-rational, fair and maximum matching. In Theorem \ref{thm:maximum}, we show that the output matching 
has the largest total trade volume among all the matchings that are individual-rational. 

\item{\textbf{Uniqueness theorems}:}  For any two fair and optimal individual-rational-uniform matchings, Theorem~\ref{thm:uniquenessTheorem} implies that their total trade volume for each order is the same. Thus, if we compare the trade volumes between an exchange's matching output with our verified program's output and for some order they do not match, then the exchange's matching is not fair and optimal individual-rational-uniform. On the contrary, if for each order, the trade volumes match, then Theorem~\ref{thm:uniquenessTheorem2} implies that the exchange's matching is also fair and optimal individual-rational-uniform (given that it already is individual-rational and uniform, which can be easily verified by checking the trade prices). Making use of these results, in Section~\ref{sec:demonstration}, we check violations automatically in real data from an exchange.

\end{itemize}

The Coq code together with the extracted OCaml and Haskell programs for all the above results is available at \cite{git:mdsa}.
Our Coq formalization consists of approximately 50 new definitions, 750 lemmas and theorems and 12000 lines of code. In the following sections, we provide definitions, procedures and proof sketches that closely follow our actual formalization.

\section{Modeling Double Sided Auctions}
\label{sec:modeling}

In a double sided auction multiple buyers and sellers place their orders to buy or sell an underlying product. The auctioneer matches these buy-sell requests  based on their \emph{limit prices}, \emph{arrival time}, and the maximum specified \emph{trade quantities}. Note that the limit prices are natural numbers when expressed in the monetary unit of the lowest denomination (like cents in USA). In our presentation, we will be working with lists (of bids, asks and transactions); For ease of readability, we will often use set-theoretic notations like $\in$, $\subseteq$, $\supseteq$, $\emptyset$ on lists whose meanings are easy to guess from the context.

\begin{definition}\textcolor{gray}{Bid}: A bid $b = (id_b, \tau_b, q_b, p_b)$ represents a buy request having four components. Here, the first two components $id_b$ and $\tau_b$ are the unique identifier and the timestamp assigned to the buy request $b$, respectively, whereas the third component $q_b$ represents the quota of $b$, the maximum quantity of the item the buyer is willing to buy. The last component $p_b$ is the limit price of the buy request, which is the price above which the buyer does not want to buy the item. 
\end{definition}

\begin{definition}\textcolor{gray}{Ask}: An ask $a = (id_a, \tau_a, q_a, p_a)$ represents a sell request having four components. Here, the first two components $id_a$ and $\tau_a$ are the unique identifier and the timestamp assigned to the sell request $a$, respectively, whereas the third component $q_a$ represents the quota of $a$, the maximum quantity of the item the seller is willing to sell. The last component $p_a$ is the limit price of the sell request, which is the price below which the seller does not want to sell the item. 
\end{definition}

We say that a bid $b \in B$ is matchable with an ask $a \in A$ if $p_a \leq p_b$.

In a double sided auction, the auctioneer is presented with duplicate-free\footnote{A list of bids or asks is duplicate-free if all the participating orders have distinct ids.} lists of buy and sell requests (lists $B$ and $A$, respectively). The auctioneer can match a bid $b \in B$ with an ask $a \in A$ only if $p_b \geq p_a$. Furthermore, the auctioneer assigns a trade price and a trade quantity to each matched bid-ask pair, which finally results in  a \emph{transaction} $m$. Therefore, we can represent a matching of demand and supply by using a list whose entries are \emph{transactions}. 

\begin{definition}\textcolor{gray}{Transaction}: A transaction $m = (b_m, a_m, q_m, p_m)$ describes a trade between the bid $b_m$ and the ask $a_m$. The next two components $q_m$ and $p_m$ are the traded quantity and the trade price, respectively. 
For ease of readability, we use the terms $p (b_m)$, $p (a_m)$, $q (b_m)$, and $q (a_m)$ for $p_{b_m}$, $p_{a_m}$, $q_{b_m}$ and $q_{a_m}$, respectively.
\end{definition}

\begin{definition}\label{def:matching}

\textcolor{gray}{(Matching M B A)}: A list of transactions $M$ is a matching between the duplicate-free lists of bids $B$ and asks $A$ if
\begin{enumerate}
\item For each transaction $m \in M$, the bid of $m$ is matchable with the ask of $m$ (i.e., $p(a_m) \leq p(b_m)$).
\item The list of bids present in $M$, denoted by $B_M$, is a subset of $B$ (i.e., $B_M \subseteq B$).
\item The list of asks present in $M$, denoted by $A_M$, is a subset of $A$ (i.e., $A_M \subseteq A$).
\item For each bid $b \in B$, the total traded volume of bid $b$ in the matching $M$, denoted by $Q(b,M)$, is not more than its maximum quantity (i.e., $\text{ for all } b \in B, Q(b, M) \leq q_b$).
\item For each ask $a \in A$, the total traded volume of ask $a$ in the matching $M$, denoted by $Q(a,M)$, is not more than its maximum quantity (i.e. $\text{ for all } a \in A, Q(a, M) \leq q_a$).
\end{enumerate}

\end{definition}

\emph{Description}. Note that there might be some bids in $B$ which are not matched to any asks in $M$  and some asks in $A$ which are not matched to any bids in $M$. 

\note{
For simplicity, with slight abuse of notation, we use $Q$ to denote total quantity of various objects which will be clear from the context. So, $Q(b,M)$ and $Q(a,M)$ represent the total quantities of the bid $b$ and the ask $a$ traded in the matching $M$, respectively. Similarly, the terms $Q(B)$ and $Q(A)$ denote the sum of the quantities of all the bids in $B$ and the sum of the quantities of all the asks in $A$, respectively. And also, for the total traded quantity in a matching $M$, we use the term $Q(M)$. However, in the Coq implementation, each of these terms are represented by different names: $\mathsf{QMb}, \mathsf{QMa}, \mathsf{QB}, \mathsf{QA} \text{ and } \mathsf{QM}$.
}

Formalization notes: We have defined Bid, Ask and Transaction as record types in Coq. We define the proposition \emph{matching\_in B A M} to be true if and only if $M$ is a matching between the list of bids $B$ and the list of asks $A$.

\subsection{Matching Demand and Supply}
Let $B_{\geq p}$ represents the list of bids in $B$ whose limit prices are at least a given number $p$. Similarly, $A_{\leq p}$ represents the list of asks in $A$ whose limit prices are at most $p$.  Therefore, the quantities $Q (B_{\geq p})$ and $Q (A_{\leq p})$ represents the total demand and the total supply of the product at the price $p$ in the market, respectively. Although, in general we cannot say much about the relationship between the total demand (i.e. $Q (B_{\geq p})$) and supply (i.e. $Q (A_{\leq p})$) at an arbitrary price $p$, we can prove the following important results about the traded quantities of the matched bid-ask pairs.

\begin{lemma}\label{lem:absumQ}
If $M$ is a matching between the list of bids $B$ and the list of asks $A$, then
\begin{align*}
Q(M) = \sum_{b \in B} Q(b,M) \le \sum_{b \in B} q_b = Q(B) \text{ and } Q(M) = \sum_{a \in A} Q(a,M) \le \sum_{a \in A} q_a = Q(A)
\end{align*}
\end{lemma}

\begin{theorem}\label{thm:boundM}
If $M$ is a matching between the list of bids $B$ and the list of asks $A$, then for all natural numbers $p$, we have
$Q(M) \leq Q(B_{\geq p}) + Q(A_{\leq p})$
\end{theorem}  

Theorem~\ref{thm:boundM} states that no matching $M$ can achieve a trade volume higher than the sum of the total demand and supply in the market at any given price.

\emph{Proof Idea.}
We first partition the matching $M$ into two lists: $M_1=\{m \in M \mid  p (b_m) \geq p\}$ and $M_2=\{m \in M \mid  p (b_m) < p\}$. Thus, $Q (M) = Q(M_1) + Q(M_2)$. 

It is easy to see that $M_1$ is a matching between $B_{\geq p}$ and $A$, and hence from Lemma~\ref{lem:absumQ}, $Q(M_1) \leq Q(B_{\geq p})$.

Next, we prove that $M_2$ is a matching between $B$ and $A_{\leq p}$. Consider a transaction $m$ from $M_2$. Since $m \in M$, $p (b_m) \ge p (a_m)$, and from the definition of $M_2$, we have $p (b_m) < p$. This implies $p (a_m) < p$, i.e., asks of $M_2$ come from $A_{\leq p}$. Hence, $M_2$ is a matching between $B$ and $A_{\leq p}$, and applying Lemma~\ref{lem:absumQ}, we have $Q(M_2) \le Q(A_{\leq p})$.

Combining, we have $Q (M) = Q(M_1) + Q(M_2) \leq Q(B_{\geq p}) + Q(A_{\leq p})$, which completes the proof of Theorem~\ref{thm:boundM}.
\hfill $\square$

Formalization notes: The formal proof of Theorem~\ref{thm:boundM} is completed by first proving the Lemmas \emph{Mbgep\_bound} ($Q(M_1) \leq Q(B_{\geq p})$) and \emph{Mbltp\_bound} ($Q(M_2) \leq Q(A_{\leq p})$) and then combining them in theorem \emph{bound\_on\_M}. These results can be found in the file 'Bound.v'.

\subsection{Individual-Rational Trades}
\label{sec:IR}
An auctioneer assigns a trade price to each matched bid-ask pair. In any matching it is desired that the trade price of a bid-ask pair lies between their limit prices. A matching which has this property is called an \emph{individual-rational (IR)} matching. 

\newcommand{\IsIR}{\mathsf{Is\_IR}}
\begin{definition}
$\IsIR(M) := \text{ for all } m \in M,\ p(b_m) \ge p_m \ge p(a_m)$.
\end{definition}

\begin{figure}[ht]
\begin{center}
\resizebox {.80\textwidth} {!} {
\begin{tikzpicture}[font=\Large]
\foreach \c/\x/\t/\y/\m in {3.3/3.9/red/69/.4, 5.05/5.2/blue/82/-.4, 5.8/6.1/green/91/-.4, 7.5/8.2/orange/112/.4, 8.9/9.5/magenta/125/.4}
{
 \draw [color=\t] (\x, 1.25) node {$]$};
 \draw [color=\t] (\x, 0) node {$]$};

 \draw [color =\t] (\x, 1.25+\m) node {{\scriptsize \y}};
 \draw [color =\t] (\x, \m) node {{\scriptsize \y}};

 \fill[color =\t] (\c,1.25) circle (0.35mm);
}

\foreach \c/\x/\t/\y/\m in {4.1/2.7/red/52/.4,4.5/4.9/blue/79/.4, 5.8/5.5/green/85/.4, 7.5/6.8/orange/98/-.4, 8.9/8.3/magenta/113/-.4}
{
 \draw [color =\t] (\x, 1.25) node {$[$};
 \draw [color =\t] (\x, 0) node {$[$};

 \draw [color =\t] (\x, 1.25+\m) node {{\scriptsize \y}};
 \draw [color =\t] (\x, \m) node {{\scriptsize \y}};
 
 \fill[color =\t] (\c, 0) circle (0.35mm);
}

\draw (2.5,0) -- (10,0);
 
\draw (2.5,1.25) -- (10,1.25);

\draw (2.25,0) node {{\scriptsize $M_2$}};
 
\draw (2.25,1.25) node {{\scriptsize $M_1$}};

\end{tikzpicture}
}
\end{center}
\caption{ The colored dots represent trade prices for matched bid-ask pairs. Matching $M_2$ is not IR but $M_1$ is IR, even though both the matchings contain exactly the same bid-ask pairs.}
\label{fig:IR}
\end{figure}
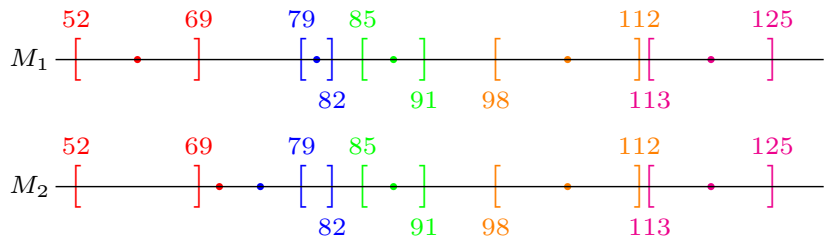

Note that any matching can be converted to individual-rational by changing the price of each transaction to lie between the limit prices of its bid and ask (See Fig~\ref{fig:IR}).

\section{Fairness in Competitive Markets}
\label{sec:fair}
A double sided auction is a competitive event, where the priority among participating traders is determined by various attributes of the orders. A bid with higher limit price is considered more \emph{competitive} compared to bids with lower limit prices. Similarly, an ask with lower limit price is considered more competitive compared to asks with higher limit prices. Ties are broken in favor of the requests that have an earlier arrival time. A matching which prioritizes more competitive traders is called a \emph{fair} matching. 
\begin{definition}[Arrow notation]
 $\stackrel{\mathbb{B}}{\uparrow} L$ denotes that the list $L$ is sorted as per the competitiveness of the bids in $L$, with the most competitive bid being on top. Similarly, $\stackrel{\mathbb{A}}{\uparrow} L$ denotes that the list $L$ is sorted as per the competitiveness of the asks in $L$, with the most competitive ask being on top. Similarly we can define $\stackrel{\mathbb{A}}{\downarrow}$ and $\stackrel{\mathbb{B}}{\downarrow}$ for sorting lists where the most competitive orders lie at the bottom.
\end{definition}

In this section, we show that there exists a procedure $\Fair$ that takes a matching $M$ between bids $B$ and asks $A$ as input and outputs a fair matching $M' = \Fair(M, B, A)$ with the same trade volume as that of $M$. To describe the $\Fair$ procedure, we will need the following definitions.

\begin{definition}
Let $M$ be a matching between bids $B$ and asks $A$. 
\begin{itemize}
\item $M$ is fair on bids if for all pairs of bids $b_1, b_2 \in B$ such that $b_1$ is more competitive than $b_2$ and $b_2$ participates in the matching $M$, then $b_1$ is fully traded in $M$ (i.e., $Q(b_1, M) = q(b_1)$).
\item Similarly,  $M$ is fair on asks if for all pairs of asks $a_1, a_2 \in A$ such that $a_1$ is more competitive than $a_2$ and $a_2$ participates in the matching $M$, then $a_1$ is fully traded in $M$ (i.e., $Q(a_1, M) = q(a_1)$).
\item $M$ is fair if it is both fair on bids and asks.
\end{itemize} 
\end{definition}

The $\Fair$ procedure works in two steps: first, it sorts the matching $M$ and the asks $A$ based on the competitiveness of the asks and then runs on them a procedure "fair on asks" $\FOA$ that outputs a matching $M'$ that is of the same volume as that of $M$ and is fair on the asks. In the second step, it sorts the resulting matching $M'$ and the bids $B$ based on the competitiveness of the bids and then runs on them a procedure fair on bids $\FOB$ that outputs a matching $M''$ that is of the same volume as that of $M'$ and is fair on the bids. The $\Fair$ procedure returns $M''$ as its output. The procedures $\FOB$ and $\FOA$, along with their correctness proofs, mirror each other and we just describe $\FOB$ below and show that $\FOB(\stackrel{\mathbb{B}}{\uparrow} M', \stackrel{\mathbb{B}}{\uparrow} B)$ outputs a fair on bids matching and has the same trade volume as that of $M'$. Furthermore, we will show that if $M'$ is fair on asks, then $\FOB(\stackrel{\mathbb{B}}{\uparrow} M', \stackrel{\mathbb{B}}{\uparrow} B)$ is fair on asks. This will immediately imply that the procedure $\Fair(M,B,A)$ outputs a fair matching with the same total trade volume as that of $M$.

\subsection{Fair on Bids}
To describe the fair on bids $\FOB$ procedure, we first need the following notation. 
\begin{definition}
Given a list $L$ and and an element $a$, $a::L$ denotes the list whose top element (head) is $a$ and the following elements (tail) are the elements of $L$ (in the same order as they appear in $L$).
\end{definition}

The $\FOB$ procedure takes sorted (based on the bids' competitiveness) lists of transactions $M$ and bids $B$. Intuitively, when all the bids are of unit quantity, we want to scan the list of transactions in $M$ from top to bottom replacing the bids therein with the bids of $B$ from top to bottom. So, in effect, in the $\FOB$ procedure we will implement this intuition apart from taking care of multiple quantity bids; and also make the procedure recursive so that we can provide a formalization friendly inductive proof of correctness.  Let $B=b::B'$ and $M=m::M'$.  In our procedure, we first pick the top bid $b$ of $B$ and the top transaction $m$ of $M$, and compare $q_b$ with $q_m$. Now we have three cases. In each of the three cases, the procedure first outputs a transaction between the bid $b$ and the ask of $m$ of quantity $\min\{q_m,q_b\}$. Case I: If $q_b = q_m$, we remove $b$ and $m$ from their respective lists and recursively solve the problem on $B'$ and $M'$. Case II: If $q_b < q_m$, we remove $b$ from the list $B$ and update $q_m $ to $q_m - q_b$ and recursively solve the problem on $B'$ and $M$. Case III: If $q_b > q_m$, we remove $m$ from the list $M$ and set a parameter $t$ to $q_m$ that we will send to the recursive call along with $M'$ and $B$. The parameter $t$ informs our recursive procedure that the top element $b$ of $B$ has effectively quantity $q_b - t$. Thus, our procedure will take three parameters: the list of transactions, the list of bids and the parameter $t$ (Note that unlike Case II ($q_b < q_m$) where the top transaction is updated, the top bid is not updated in Case III ($q_b > q_m$). This is done for technical reasons: Later we need to prove that the set of bids of $\FOB$ is a subset of $B$, and at the same time we have to ensure that the total traded quantity of the bid $b$ in the matching outputted by $\FOB$ remains below its maximum quantity $q_b$, as required by the matching property. This would not be possible to do if we updated $B$ and hence we take this approach of keeping the total traded quantity of the top bid $b$ in a separate argument: $t$ of $f$.). Keeping this description in mind, we now formally define the procedure $\FOB$.

\begin{definition} \textcolor{gray}{Fair On Bid ($\FOB$)}.
\begin{flalign*}
&\FOB(M,B)= f(M,B,0) \\
	&\text{where } f(M,B,t) =  \\
	 & \begin{cases}
	 	nil  &\text{\hspace{-1cm} if $M = nil$ or $B = nil$} \\
		(b,a_m,q_m,p_m)::f(M',B',0) & \text{if $q_m=q_b - t$} \\
		(b,a_m,q_m,p_m)::f(M',b::B',t+q_m) & \text{if $q_m < q_b - t$} \\
		(b,a_m,q_b-t,p_m)::f((b_m,a_m,q_m-(q_b-t),p_m)::M',B',0) & \text{if $q_m > q_b - t$} 
	\end{cases}
\end{flalign*}
where $M=m::M'$ when $M \neq nil$ and $B = b::B'$ when $B \neq nil$.
\end{definition}

\emph{Description}. In non-trivial cases (i.e. $M \neq nil$ and $B \neq nil$) the function call $\FOB (M, B)$ reduces to $f(M,B,0)$.  


\begin{theorem}
\label{thm:FOB}
Let $M$ be a matching between bids $B$ and asks $A$ where the lists $M$ and $B$ are sorted in the descending order of the competitiveness of their bids (i.e., the most competitive bid and the transaction with the most competitive bid are on top of their respective lists). Let $M_\beta = \FOB (M, B)$, then

\begin{enumerate}[label=(\alph*)]
\item $M_\beta$ is a matching between bids $B$ and asks $A$.
\item For each ask $a \in A$, the total traded quantity of $a$ in $M$ is same as the total traded quantity of $a$ in $M_\beta$ (i.e., $Q(a,M) = Q(a,M_\beta)$). As a corollary, we get that if $M$ is fair on asks, then $M_\beta$ is also fair on asks.
\item The total traded quantity of $M$ is equal to the total traded quantity of $M_\beta$ (i.e., $Q(M) = Q(M_\beta)$).
\item The matching $M_\beta$ is fair on bids.
\end{enumerate}

\end{theorem}

\emph{Proof Outline}. Here, we briefly describe certain aspects of the proof; more details can be found in Appendix~\ref{fairappendix} and for the complete formalization see \cite{git:mdsa}. Note that in each of the recursive calls in $f$, either the size of the first argument $|M|$ decreases or the size of the second argument $|B|$ decreases. Therefore, we prove the above statements using (well founded) induction on the sum $(|M|+|B|)$. Proof of (a) and (b) is done using induction and case analysis. The proof of (c) follows by combining Lemma~\ref{lem:absumQ} with (b). We focus on the proof of (d) below.

Let bid $b$ be the top element of the bids $B$ and $B = b::B'$. First, we prove two general results: 
$$\text{For all } t, \text{ if }Q(M)  \ge q_b - t, \text{ then } Q(f(M,b::B',t), b) = q_{b} - t,$$ 
which states that if the total trade volume of the matching $M$ is at least $q_b - t$, then in the matching $f(M,b::B',t)$ the top bid $b$ has trade quantity $q_b - t$. The proof of this can be done using induction on the size of $M$. Intuitively, $f$ tries to match as much quantity of the top bid $b$ with the top transaction in $M$. When the call $f(M,b::B',t)$  is made, the top bid $b$ already has $t$ traded quantity and $q_b - t$ of its quantity remains untraded. If the quantity of the top transaction $m$ of $M$ is at least $q_b - t$, then we are done. Otherwise, $f$ matches $q_m$ quantities of $b$ and recursively calls $f$ on a smaller list and then we will be done by applying the induction hypothesis.

Now, we state the second general result. 
$$\text{For all } t,\text{ if distinct bids }b, b'\text{ belong to the bids of }f(M,b::B',t),\text{ then }Q(M) \ge q_{b} - t.$$ 
This result can be proved, like the previous result, using induction on the sum ($|M|+|B|$); see~\cite{git:mdsa} for details. Intuitively, since $b'$ is matched by $f(M,b::B',t)$ (in particular $b' \in B'$), then $f(M,b::B',t)$ will completely match $b$ (which has at least $q_b - t$ quantity remaining untraded) before it matches even a single quantity of $b'$. 

Now using the above general results, we prove (d). We need  to show the following: for all $b_1, b_2 \in B$, if $b_1$ is more competitive than $b_2$ and  $Q(\FOB(M, B), b_2) \ge 1$, then $Q(\FOB(M,B), b_1) = q_{b_1}$, i.e., if the bid $b_2$ participates in the matching $M$ then the bid $b_1$ is fully traded in $M$. Fix $b_1,b_2 \in B$ such that $b_1$ is more competitive than $b_2$ and $Q(\FOB(M, B), b_2) \ge 1$. Note that the bid $b_2$ cannot be equal to the bid $b$ since bids $B$ are sorted. Now we analyze three possible cases: $b_1 \neq b$, $b_1 = b$ and $Q(M) \ge q_b$, and $b_1 = b$ and $Q(M) < q_b$.
\begin{itemize}
\item In the case when $b_1 \neq b$, we consider the recursive call where $b_1$ is the top bid in the argument for the first time. In this recursive call the list of bids is smaller than $B$ since the bid $b$ must be fully traded before. Then, we are immediately done by applying the induction hypothesis.

\item In the case $b_1 = b$ and $Q(M)  \ge q_b $, in the matching $\FOB(M,b::B') = f(M,b::B',0)$ the top bid $b$ has total trade volume $q_b - 0 = q_b$ from the first general result invoked with $t=0$, and hence $b_1 = b$ is fully traded.

\item In the case $b_1 = b$ and  $Q(M)  < q_b $,  we arrive at the contradiction $Q(M)\ge q_b$ by invoking the second general result with $t=0$, $b = b_1$, $b' = b_2$ and $\FOB(M,b::B') = f(M,b::B',0)$.
\end{itemize} \hfill $\square$

Similar to the procedure $\FOB$, we have a procedure $\FOA$, that produces a fair matching on asks (see \cite{git:mdsa}). Combining the $\FOA$ and $\FOB$ procedures, we have the following definition of the $\Fair$ procedure.

\begin{definition} \label{def:fair} 
$\Fair(M ,B ,A) =  \FOB(\stackrel{\mathbb{B}}{\uparrow} \FOA (\stackrel{\mathbb{A}}{\uparrow} M , \stackrel{\mathbb{A}}{\uparrow} A), \stackrel{\mathbb{B}}{\uparrow} B)$.
\end{definition}

We conclude this section by formally summarizing the main fairness result.

\begin{theorem} \label{thm:existsFair}
If $M$ is a matching on the list of bids $B$ and the list of asks $A$, then the matching $M' = \Fair (M ,B ,A)$ on $B$ and $A$ is a fair matching such that $Q(M) = Q(M')$.
\end{theorem}

Formalization notes: The procedure $\FOB$ and $\FOA$ are implemented in Coq using the Equations plugin which is helpful to write functions involving well-founded recursion \cite{equations}. The proof of Theorem~\ref{thm:existsFair} is quite extensive and done in several parts. First we prove all the parts of Theorem~\ref{thm:FOB} in the file 'mFair\_Bid.v'. We prove similar theorems for the procedure $\FOA$ in 'mFair\_Ask.v' file. Later all the results are combined in the file 'MQFair.v' and the above theorem is proved as \emph{exists\_fair\_matching}.

\section{Uniform Price Matchings in Financial Markets}
\label{sec:uniform}
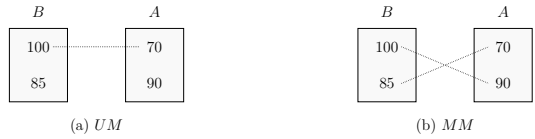
\begin{figure}[ht]
\begin{center}

\resizebox {.5\textwidth} {!} {

\begin{tikzpicture}[font=\Large]

 \draw [fill=gray!5] rectangle (22/14,2);

 \draw (.5*22/14, .5) node {$85$};

 \draw (.5*22/14, 1.5) node {$100$};

 \draw [fill=gray!5] (2*22/14,0) --(2*22/14,2)-- (3*22/14,2)-- (3*22/14,0)--cycle;

 \draw (2.5*22/14, .5) node {$90$};

 \draw (2.5*22/14, 1.5) node {$70$};

 \draw [fill=gray!5] (6*22/14,0) --(6*22/14,2)-- (7*22/14,2)-- (7*22/14,0)--cycle;

 \draw (6.5*22/14, .5) node {$85$};

 \draw (6.5*22/14, 1.5) node {$100$};

 \draw [fill=gray!5] (8*22/14,0) --(8*22/14,2)-- (9*22/14,2)-- (9*22/14,0)--cycle;

 \draw (8.5*22/14, .5) node {$90$};

 \draw (8.5*22/14, 1.5) node {$70$};

 \path (.5*22/14, 2.2) node[above] {$B$};

 \path (2.5*22/14, 2.2) node[above] {$A$};
sex
 \path (6.5*22/14, 2.2) node[above] {$B$};

 \path (8.5*22/14, 2.2) node[above] {$A$};

  \draw (1.5*22/14, -1) node[above] {(a) $UM$};
  \draw (7.5*22/14, -1) node[above] {(b) $MM$};


  \draw [densely dotted] (0.75*22/14,1.5) -- (2.25*22/14,1.5);

  \draw [densely dotted] (6.75*22/14,1.5) -- (8.25*22/14,0.5);

  \draw [densely dotted] (6.75*22/14,0.5) -- (8.25*22/14,1.5);

\end{tikzpicture}
}
\end{center}
\caption{Both the bids and the asks have quantity one. The only individually rational matching of size two is not uniform.}
\label{fig:mmum}

\end{figure}

Liquidity in a market is a measure of how quickly one can trade in that market and maximizing the total trade volume helps increase liquidity. 
However, to maximize the total trade volume sometimes we have to accept different trade prices to the matched bid-ask pairs (Fig ~\ref{fig:mmum}). 

Assigning different trade prices for the same product in the same market simultaneously, might lead to dissatisfaction among some traders. As stated in the introduction, in the financial markets, the matching should be fair and optimal individual-rational-uniform. In this section, we describe the $\UM$ process that takes as input a list of bids and a list of asks and produces a fair and optimal individual-rational-uniform matching that can be directly applied in the financial markets for conducting double sided auctions. We present a novel proof of optimality of the $\UM$ process.

Before we describe the $\UM$ process, we first give some intuition. Observe that in any individual-rational and uniform matching $M$ all the buyers are matched at a single price $p$ and the price $p$ lies between the limit prices of all the matched bid-ask pairs. This means all the matched bids' limit prices are at least $p$ and all the matched asks' limit prices are at most $p$. In the special case when all the orders are of unit quantity, the matching can be visualized as a fully nested balanced parenthesis (for example, [[[[ ]]]]) where each bid is represented by a closed parenthesis `]' and each ask as an open parenthesis `[' (See Figure~\ref{fig:IR}).

Now, we describe the $\UM$ process. We recursively pair the most competitive available bid with the most competitive available ask, if they are matchable. The trade quantity for each matched bid-ask pair is the minimum of the remaining quantities of the respective bid and the ask. The trade price assigned to each pair is the price of the ask in that pair\footnote{Observe that any value in the interval of the limit prices of the matched bid-ask pair can be assigned as the trade price and it will not affect any analysis done in this work.}. We terminate the process once there are no more matchable bid-ask pairs remaining. At the end of the process, to produce a uniform matching we have to assign a single trade price to all the matched bid-ask pairs which we choose to be the trade price of the last matched bid-ask pair (which also keeps the individual-rational property intact).

Keeping this description in mind, we now formally define the $\UM$ process using recursion.

\begin{definition} \textcolor{gray}{Uniform Matching ($\UM$)}.
\begin{flalign*}
& \UM(B, A) = \mathsf{Replace\_prices}(f_u(\stackrel{\mathbb{B}}{\uparrow} B, \stackrel{\mathbb{A}}{\uparrow} A,0,0) , \mathsf{Last\_trade\_price} (f_u(\stackrel{\mathbb{B}}{\uparrow} B, \stackrel{\mathbb{A}}{\uparrow} A,0,0))) \\
&\text{where } f_u(B,A,t_b,t_a) = \\
	 &\begin{cases}
	 	nil  &\text{if $B = nil$ or $A = nil$ or $p_b < p_a$} \\
		(b,a,q_b - t_b,p_a)::f_u(B',A',0,0) & \text{if $q_a - t_a=q_b - t_b \text{ and } p_b \ge p_a$} \\
		(b,a,q_b - t_b,p_a)::f_u(B',a::A',0,t_a + q_b - t_b) & \text{if $q_a - t_a > q_b - t_b \text{ and } p_b \ge p_a$} \\
		(b,a,q_a - t_a,p_a)::f_u(b::B',A',t_b + q_a - t_a,0) & \text{if $q_a - t_a < q_b - t_b \text{ and } p_b \ge p_a $}		
	\end{cases}
\end{flalign*}
where $B=b::B'$ when $B \neq nil$ and $A = a::A'$ when $A \neq nil$.
\end{definition}

\begin{description}
\item \emph{Description}. Observe that, similar to the parameter $t$ in the $\FOB$ process, we have two parameters $t_b$ and $t_a$ that inform the recursive procedure $f_u$ that the top bid $b$ and the top ask $a$ have effective quantities $q_b - t_b$ and $q_a - t_a$, respectively. In each recursive call, the process $f_u$ outputs a transaction (top bid $b$, top ask $a$, quantity $\min\{q_b - t_b, q_a - t_a\}$, price $p_a$). The process $f_u$ terminates when the top bid is not matchable with the top ask. 
\item \emph{Remark 1}. It is easy to see that $\UM$ outputs a uniform matching: Once the $f_u$ process terminates, $\mathsf{Last\_trade\_price}$ computes the trade price of the last transaction in the output of $f_u$ and $\mathsf{Replace\_prices}$ replaces the trade prices of each transaction of the output of $f_u$ with the trade price of the last transaction of the output, thus ensuring $\UM$ produces a uniform matching. Also, notice that the process 
$\mathsf{Replace\_prices}$ does not alter any other information of the output of $f_u$ apart from the trade prices (we will later use this fact in the proof of optimality of $\UM$). 
\item \emph{Remark 2}. It is easy to see that $\UM$ outputs an individual-rational matching: the trade price of a transaction $m$ outputted by a recursive call of $f_u$ is between the limit prices of the bid and the ask of $m$. Later these prices are altered by $\mathsf{Replace\_prices}$, but the individual-rational property is not lost; the trade price of $m$ is also between the limit prices of the transactions of all the previous calls as the bids and the asks are sorted by their competitiveness, and $\mathsf{Replace\_prices}$ replaces all the trade prices with the trade price of the last transaction.
\end{description}

Now, we discuss the optimality result of the $\UM$ process. Throughout this discussion, WLOG, all lists of bids and asks will be sorted by their competitiveness. We make use of the following notation.

\begin{definition}
Given a matching $M$, a bid $b$ and an ask $a$, we use $Q(a \leftrightarrow b, M)$ to denote the total traded quantity between the bid $b$ and the ask $a$ in the matching $M$.
\end{definition}

Next, we state the main result of this section.

\begin{theorem} \label{thm:UMopt}
Given a list of bids $B$ and a list of asks $A$, let $M_U = \UM (B, A)$ and let $M$ be an arbitrary individual-rational and uniform matching between $B$ and $A$. Then, $Q(M_U) \ge Q(M)$. In other words, $\UM$ outputs an optimal individual-rational-uniform matching.
\end{theorem}

To prove the above theorem, we need the following lemma.

\begin{lemma} \label{lem:surgery1}
If $M$ is an individual-rational and uniform matching between the lists of bids $B = b::B'$ and asks $A = a::A'$ such that $Q(M) \ge \min\{q_b,q_a\}$, then there exists another individual-rational and uniform matching $M'$ between the same lists of bids $B$ and asks $A$ such that $Q(M) = Q(M')$ and $Q(a \leftrightarrow b, M') = \min\{q_b,q_a\}$.
\end{lemma}

Assuming this lemma, we will first prove Theorem~\ref{thm:UMopt} and then later prove the lemma.

\emph{Proof of Theorem~\ref{thm:UMopt}}.
Note that $f_u(B,A,0,0)$ is a specific instance of $f_u(B,A,t_b,t_a)$. So in order to apply the induction hypothesis, we sensitize the theorem statement to incorporate arbitrary values of $t_a$ and $t_b$. Also, as indicated earlier, the $\mathsf{Replace\_prices}$ function does not alter the total trade quantity of the output of the $f_u$, thus $Q(f_u(B,A,0,0)) = Q(\UM(B,A))$. Consequently, showing the following suffices.

\begin{itemize}
\item [($\ast$)] Fix an arbitrary list of bids $B = b::B'$ and an arbitrary list of asks $A = a::A'$.  Fix arbitrarily $t_b < q_b$ and $t_a < q_a$. Let $b'$ be the bid obtained from the bid $b$ by reducing its quantity to $q_b - t_b$. Similarly, let $a'$ be the ask obtained from the ask $a$ by reducing its quantity to $q_a - t_a$. We will show: for all individual-rational and uniform matchings $M$ between $(b'::B')$ and $(a'::A')$, $Q(f_u(B,A,t_b,t_a)) \ge Q(M)$.
\end{itemize}

Clearly, setting $t_b = t_a = 0$ in the above statement ($\ast$) gives us Theorem~\ref{thm:UMopt}. 

We prove the above statement using induction on the sum $(|B| + |A|)$. We consider two cases: $p(b')<p(a')$ and $p(b')\ge p(a')$. 


In the first case, when $p(b') < p(a')$, since the most competitive bid in $b'::B'$ is not matchable with the most competitive ask in $a'::A'$, any matching between $b'::B'$ and $a'::A'$ is empty. Thus, $Q(M) = 0$, and we are done.

In the second case, when $p(b')\ge p(a')$, if the total trade quantity of $M$ is less than the quantity of the transaction created by $f_u$ in the first recursive call (i.e., $Q(M) < \min\{q_b - t_b, q_a - t_a\} \le Q(f_u(B,A,t_b,t_a)$), then we are done. In the case when the total traded quantity of $M$ is more than the quantity of the transaction created by $f_u$ in the first recursive call (i.e., $Q(M) \ge \min\{q_b - t_b, q_a - t_a\}$), we apply Lemma~\ref{lem:surgery1} and get another individual-rational and uniform matching $M'$ such that the total volume of $M'$ is equal to the total volume of $M$ and the total traded quantity between the bid $b'$ and the ask $a'$ in $M'$ is equal to $\min\{q_{b'} = q_b - t_b, q_{a'} = q_a - t_a\}$. Now since we have $M'$ such that $Q(M) = Q(M')$, proving the following suffices.

$$Q(f_u(B, A,t_b,t_a)) \ge Q(M') \quad\quad\quad\quad\quad\quad (\ast\ast),$$ 

where $M'$ is an individual-rational and uniform matching between the list of bids $b'::B'$ and $a'::A'$ such that $Q(a' \leftrightarrow b', M') = \min\{q_b - t_b,q_a - t_a\}$. We define the matching $M_0 \subseteq M$ as follows: we remove all transactions between $b'$ and $a'$ (of total quantity $Q(a' \leftrightarrow b', M')$) from $M'$ to get $M_0$. We have ($\dagger$): $Q(M') = \min\{q_a - t_a, q_b - t_b\} + Q(M_0)$. Also, note that $M_0$ is individual-rational and uniform (since $M' \supseteq M_0$ is individual-rational and uniform).

Now we argue the proof of $(\ast\ast)$ in each of the three recursive branches of the function $f_u$ corresponding to $p(b) \ge p(a)$.

\begin{itemize}
\item Case: $q_a - t_a= q_b - t_b$. In this case $M_0$ is a matching between $B'$ and $A'$. Since $(|B|+|A|) > (|B'|+|A'|)$, we can apply the induction hypothesis to get $Q(f_u(B',A',0,0)) \ge Q(M_0)$. Now, applying the definition of $f_u$ we get,
\begin{align*}
Q(f_u(B,A,t_b,t_a)) &= \min\{q_a - t_a, q_b - t_b\} + Q(f_u(B',A',0,0)) \\
&\stackrel{\text{I.H.}}{\ge} \min\{q_a - t_a, q_b - t_b\} + Q(M_0) \stackrel{(\dagger)}{=} Q(M').
\end{align*}
 
\item Case: $q_a - t_a > q_b - t_b$. In this case $M_0$ is a matching between $B'$ and $\hat{a}::A'$ (where $q_{\hat{a}} = q_a - t_a - (q_b - t_b) \le q_a$). 
Since $(|B|+|A|) > (|B'|+|\hat{a}::A'|)$, we can apply the induction hypothesis when $B' \neq \emptyset$ to get $Q(f_u(B',A,0,t_a + (q_b - t_b))) \ge Q(M_0)$. When $B' = \emptyset$, then $Q(f_u(B',A,0,t_a + (q_b - t_b))) \ge Q(M_0)$ holds trivially as both the sides of the inequality are zeros. Now, applying the definition of $f_u$ we get, 
\begin{align*}
Q(f_u(B,A,t_b,t_a)) &= (q_b - t_b) + Q(f_u(B',A,0,t_a + (q_b - t_b))) \\
&\stackrel{\text{I.H.}}{\ge} (q_b - t_b) + Q(M_0)  \stackrel{(\dagger)}{=} Q(M').
\end{align*}

\item Case: $q_a - t_a < q_b - t_b$. This is symmetric to the previous case and the proof follows similarly. \hfill $\square$
\end{itemize} 

Having finished the proof of the main result, we now discuss the proof of the lemma that we assumed.

\emph{Main proof idea of Lemma~\ref{lem:surgery1}}. 
Given an individual-rational and uniform matching $M$ with $Q(M) \ge \min\{q_b,q_a\}$ between the list of bids $B = b::B'$ and the list of asks $A = a::A'$, we need to show existence of an individual-rational and uniform matching $M'$ such that $Q(M') = Q(M)$ and the total trade quantity between the bid $b$ and ask $a$ in $M'$ is $\min\{q_b,q_a\}$. We do the following surgery on $M$ in two steps to obtain the desired $M'$. 

Step 1: We first modify $M$ to ensure that bid $b$ and ask $a$ each has at least $\min\{q_b,q_a\}$ total trades in $M$ (not necessarily between each other). This is accomplished by running the $\Fair$ procedure on $M$ that outputs a matching which prefers the most competitive orders ($b$ and $a$) over any other orders. Since $Q(M) \ge \min\{q_b,q_a\}$, we get that $\Fair(M,B,A)$ has at least $\min\{q_b,q_a\}$ trades for each of $b$ and $a$. Note that $\Fair$ does not change the total trade quantity or affect the individual-rational and uniform properties of $M$. Set $M \leftarrow \Fair(M,B,A)$.

Step 2: In this step, we modify $M$ to ensure that the bid $b$ and ask $a$ have $\min\{q_b,q_a\}$ quantity trade between them. Note that in $M$ individually both $b$ and $a$ have at least $\min\{q_b,q_a\}$ total trade quantity. We will inductively transfer trades of $b$ and $a$ that are not between them to the transaction between $b$ and $a$, a unit quantity at a time, till they have $\min\{q_b,q_a\}$ quantity trade between them. To better understand this, consider the case when $b$ and $a$ have zero trade quantity between them. Let us say there is a transaction between $b$ and $a_1$ of quantity $q_1$ and a transaction between $a$ and $b_1$ of quantity $q_2$. We remove these two transactions and replace it with the following four transactions (see Figure~\ref{fig:surgery}) that keeps the matching trade volume intact: (1) transaction between $b$ and $a_1$ of quantity $q_1 -1$, (2) transaction between $a$ and $b_1$ of quantity $q_2 - 1$, (3) transaction between $b_1$ and $a_1$ of quantity one and (4) transaction between $b$ and $a$ of quantity one. Recall, in a individual-rational and uniform matching with price $p$, the limit price of each bid is at least $p$ and the limit price of each ask is at most $p$, implying any bid and ask participating in the matching are matchable. Thus, doing such a replacement surgery is legal and does not affect the individual-rational and uniform properties, and we obtain the desired $M'$ by repeatedly doing this surgery.

\begin{figure}[ht]
\begin{center}

\resizebox {.7\textwidth} {!} {

\begin{tikzpicture}[font=\Large]

 \draw [fill=gray!5] (0,0) -- (4,0) -- (4,3) -- (0,3)--cycle;
 \fill [fill=yellow!70] (0,.5) -- (1,.5) -- (1,1) -- (0,1)--cycle;
 \fill [fill=red!20] (0,.5+1.8) -- (1,.5+1.8) -- (1,1+1.8) -- (0,1+1.8)--cycle;
 \fill [fill=blue!20] (1,.5) -- (2,.5) -- (2,1) -- (1,1)--cycle;
 \fill [fill=green!20] (1,.5+1.8) -- (2,.5+1.8) -- (2,1+1.8) -- (1,1+1.8)--cycle;
 \draw (2, 2) node[below] {$M$};
 \draw (.5, .75+1.8) node {$b$};
 \draw (1.5, .75) node {$a$};
 \draw (1.5, .75+1.8) node {$a_1$};
 \draw (.5, .75) node {$b_1$};

 \fill [fill=white] (2,.5) -- (3,.5) -- (3,1) -- (2,1)--cycle;
 \draw (2.5, .75) node {\small{$q_2$}};

 \fill [fill=white] (3,.5) -- (4,.5) -- (4,1) -- (3,1)--cycle;
 \draw (3.5, .75) node {\small{$p$}};
 \draw [dotted] (3,.5) -- (3,1);

 \fill [fill=white] (2,.5+1.8) -- (3,.5+1.8) -- (3,1+1.8) -- (2,1+1.8)--cycle;
 \draw (2.5, .75+1.8) node {\small{$q_1$}};

 \fill [fill=white] (3,.5+1.8) -- (4,.5+1.8) -- (4,1+1.8) -- (3,1+1.8)--cycle;
 \draw (3.5, .75+1.8) node {\small{$p$}};
 \draw [dotted] (3,.5+1.8) -- (3,1+1.8);
 \draw (4,0) -- (4,3) --cycle;
 \draw (0,0) -- (0,3) --cycle;

 \draw (-.5, .75+1.8) node {$m_1$};
 \draw (-.5, .75) node {$m_2$};

 \draw [fill=gray!5,xshift=2.5in] (0,0) -- (4,0) -- (4,4) -- (0,4)--cycle;
 \fill [fill=yellow!70,xshift=2.5in] (0,.5) -- (1,.5) -- (1,1) -- (0,1)--cycle;
 \fill [fill=red!20,xshift=2.5in] (0,.5+1.8) -- (1,.5+1.8) -- (1,1+1.8) -- (0,1+1.8)--cycle;
 \fill [fill=blue!20,xshift=2.5in] (1,.5) -- (2,.5) -- (2,1) -- (1,1)--cycle;
 \fill [fill=green!20,xshift=2.5in] (1,.5+1.8) -- (2,.5+1.8) -- (2,1+1.8) -- (1,1+1.8)--cycle;
 \draw (2, 2) node[below,xshift=2.5in] {$M'$};
 \draw (.5, .75+1.8) node[xshift=2.5in] {$b$};
 \draw (1.5, .75) node[xshift=2.5in] {$a$};
 \draw (1.5, .75+1.8) node[xshift=2.5in] {$a_1$};
 \draw (.5, .75) node[xshift=2.5in] {$b_1$};

 \fill [fill=white,xshift=2.5in] (2,.5) -- (3,.5) -- (3,1) -- (2,1)--cycle;
 \draw (2.5, .75) node[xshift=2.5in] {\small{$q_2 - 1$}};

 \fill [fill=white,xshift=2.5in] (3,.5) -- (4,.5) -- (4,1) -- (3,1)--cycle;
 \draw (3.5, .75) node[xshift=2.5in] {\small{$p$}};
 \draw [dotted,xshift=2.5in] (3,.5) -- (3,1);

 \fill [fill=white,xshift=2.5in] (2,.5+1.8) -- (3,.5+1.8) -- (3,1+1.8) -- (2,1+1.8)--cycle;
 \draw (2.5, .75+1.8) node[xshift=2.5in] {\small{$q_1 - 1$}};

 \fill [fill=white,xshift=2.5in] (3,.5+1.8) -- (4,.5+1.8) -- (4,1+1.8) -- (3,1+1.8)--cycle;
 \draw (3.5, .75+1.8) node[xshift=2.5in] {\small{$p$}};
 \draw [dotted,xshift=2.5in] (3,.5+1.8) -- (3,1+1.8);
 \draw [xshift=2.5in] (4,0) -- (4,4) --cycle;
 \draw [xshift=2.5in] (0,0) -- (0,4) --cycle;

 \fill [fill=yellow!70,xshift=2.5in] (0,.5+2.5) -- (1,.5+2.5) -- (1,1+2.5) -- (0,1+2.5)--cycle;
 \fill [fill=red!20,xshift=2.5in] (0,.5+3) -- (1,.5+3) -- (1,1+3) -- (0,1+3)--cycle;
 \fill [fill=green!20,xshift=2.5in] (1,.5+2.5) -- (2,.5+2.5) -- (2,1+2.5) -- (1,1+2.5)--cycle;
 \fill [fill=blue!20,xshift=2.5in] (1,.5+3) -- (2,.5+3) -- (2,1+3) -- (1,1+3)--cycle;
 \fill [fill=white,xshift=2.5in] (2,.5+2.5) -- (3,.5+2.5) -- (3,1+2.5) -- (2,1+2.5)--cycle;
 \fill [fill=white,xshift=2.5in] (3,.5+2.5) -- (4,.5+2.5) -- (4,1+2.5) -- (3,1+2.5)--cycle;
 \fill [fill=white,xshift=2.5in] (2,.5+3) -- (3,.5+3) -- (3,1+3) -- (2,1+3)--cycle;
 \fill [fill=white,xshift=2.5in] (3,.5+3) -- (4,.5+3) -- (4,1+3) -- (3,1+3)--cycle;
 \draw [xshift=2.5in] (0,0) -- (4,0) -- (4,4) -- (0,4)--cycle;
 \draw [dotted,xshift=2.5in] (0,3) -- (4,3);
 \draw [dotted,xshift=2.5in] (0,3.5) -- (4,3.5);

 \draw [xshift=2.5in] (.5, 3.25) node {$b_1$};
 \draw [xshift=2.5in] (1.5, 3.25) node {$a_1$};
 \draw [xshift=2.5in] (2.5, 3.25) node {$1$};
 \draw [xshift=2.5in] (3.5, 3.25) node {\small{$p$}};

 \draw [xshift=2.5in] (.5, 3.75) node {$b$};
 \draw [xshift=2.5in] (1.5, 3.75) node {$a$};
 \draw [xshift=2.5in] (2.5, 3.75) node {$1$};
 \draw [xshift=2.5in] (3.5, 3.75) node {\small{$p$}};

 \draw [dotted,xshift=2.5in] (3,3) -- (3,4);

 \draw [xshift=2.5in] (-.5, .75+1.8) node {$m_1'$};
 \draw [xshift=2.5in] (-.5, .75) node {$m_2'$};
 \draw (5, 1.5) node {{$\Rightarrow$}};
\end{tikzpicture}
}
\end{center}
\caption{
}
\label{fig:surgery}

\end{figure}

The proof that $\UM$ produces a fair matching follows from inducting on the sum $(|A|+|B|)$ and the fact that $B$ and $A$ are sorted by competitiveness of the participating bids and asks. The argument is similar to the correctness proof of $\Fair$ that we saw before. From the discussion above, the next theorem follows immediately.

\begin{theorem} \label{thm:um}
For a given list of bids $B$ and the list of asks $A$, $M = \UM(B ,A)$ is a fair and optimal individual-rational-uniform matching on $B$ and $A$.
\end{theorem}

Formalization notes: The formalized proof of the above theorem is done by first proving Lemma~\ref{lem:surgery1} (\emph{exists\_opt\_k}) using induction on the gap $k=\min\{q_b,q_a\} - Q(a \leftrightarrow b, M)$. From this lemma, we get another matching $M'$ such that $Q(a \leftrightarrow b, M) = \min\{q_b,q_a\}$. The matching $M'$ is altered to $M_0$ (as described in the proof of Theorem~\ref{thm:UMopt} above) by removing all the transactions between the bid $b$ and the ask $a$. We prove that the altered list $M_0$ is is a matching between the reduced lists of bids and asks. All the results related to $M_0$ are in the file 'MachingAlter.v'. Finally, combining all these we prove the main Theorem~\ref{thm:um} as \emph{'UM\_main'}.

\section{A Maximum Matching Mechanism} 
\label{sec:maximum}
In the previous section, we indicated that to achieve maximum trade volume matching we sometimes have to assign different trade prices to the matched bid-ask pairs. An individual rational matching with maximum trade volume is called a maximum matching. In this section we describe a process $\MM$, that takes a list of bids and a list of asks and outputs a fair, individual-rational and maximum matching. 

The $\MM$ procedure roughly works as follows. In step one, the $\MM$ procedure repeatedly pairs the most competitive bid $b$ with the least competitive matchable ask $a$ and outputs a transaction $(b,a,\min\{q_b,q_a\},p_a)$ and decreases the quantities of $b$ and $a$ by $\min\{q_b,q_a\}$. In step two, the $\MM$ procedure applies the $\Fair$ procedure on the output of step one. 

The detailed $\MM$ procedure and the proof of its correctness are similar to that of the $\UM$ procedure in spirit. In the proof of optimality, we need to prove a lemma similar to Lemma~\ref{lem:surgery1} which states that a given arbitrary individual-rational matching $M$ of sufficiently large trade volume can be altered to obtain a matching $M'$ of the same total trade volume such that the total trade quantity between the most competitive bid and the corresponding least competitive matchable ask in $M'$ is the minimum of their respective quantities. The proof of this requires more surgeries as compared to that in the  proof of Lemma~\ref{lem:surgery1}. Besides this deviation all other arguments of the proof of optimality of $\MM$ are similar to that of $\UM$ with minor variations.

The proof of $\MM$ producing an individual-rational matching is trivial and the proof that it produces a fair matching follows from the fact that $\MM$ applies the $\Fair$ procedure before it outputs a final matching. We now state the main theorem of this section.

\begin{theorem} \label{thm:maximum}
For a given list of bids $B$ and a list of asks $A$, $\MM(B,A)$ is a fair, individual-rational and maximum trade volume matching between $B$ and $A$.
\end{theorem}

The proof of the above theorem and discussion around the $\MM$ procedure is moved to Appendix~\ref{sec:maxappendix}. 

Formalization notes: All the formalization details can be found in \cite{git:mdsa}.

\section{Uniqueness Theorem} \label{sec:uniqueness}
In this section, we establish certain theorems that enable us to automatically check for violations in an exchange matching algorithm by comparing its output with the output of our certified program. Detailed proofs are available in the Coq formalization~\cite{git:mdsa}.

Ideally, we would have wanted a theorem that the properties (fair and optimal individual-rational-uniform) imply a unique matching. Such a theorem would enable us to automatically compare a matching produced by an exchange with a matching produced by our certified program to find violations of these properties in the matching produced by the exchange. Unfortunately, such a theorem is not possible; there exists two different matchings $M_1$ and $M_2$ on the same list of bids $B$  and asks $A$, where both are fair and optimal individual-rational-uniform: $M_1 = \{(b_1,a_1,1,p), (b_2,a_2,2,p)\}$ and $M_2 = \{(b_1,a_2,1,p), (b_2,a_2,1,p), (b_2,a_1,1,p)\}$ on bids $B = \{b_1=(*,*, 1, p), b_2 = (*,*,2, p)\}$ and asks $A = \{a_1=(*,*,  1, p), a_2 = (*,*, 2, p)\}$ for some arbitrary price $p$, timestamps and ids. Note that fairness does not require the most competitive bid to be paired with the most competitive ask. For example, assuming $a_1$ has a lower timestamp than $a_2$ and $b_1$ has a lower timestamp than $b_2$ in the above example, $a_1$ and $b_1$ are not matched in the matching $M_2$,  which is a fair matching.
Nonetheless,  we can show that given a list of bids $B$ and a list of asks $A$, all matchings that are fair and individual-rational-uniform, must have the same trade volume for each trader. This still allows us to automatically check for violations of the properties in an exchange, by comparing the trades of each trader produced by the exchange against that produced by our certified program.

We have the following lemma which formulates this uniqueness relation on the matchings.

\begin{theorem} \label{thm:uniquenessLemma}
	Let $M_1$ and $M_2$ be two fair matchings on the list of bids $B$ and the list of asks $A$ such that $Q(M_1)=Q(M_2)$, then for each order $\omega$, the total traded quantity of $\omega$ in $M_1$ is equal to the total traded quantity of $\omega$ in $M_2$.
\end{theorem}
\emph{Proof Idea}. We now prove the above theorem by using Lemma~\ref{lem:absumQ} and deriving a contradiction.
Let $M_1$ and $M_2$ be fair matchings such that $Q(M_1)=Q(M_2)$. Let $b$ be a buyer whose total trade quantity in $M_1$ is different (WLOG, more) from his total trade quantity in $M_2$. It is easy to show that there exists another buyer $b'$ such that her total traded quantity in $M_1$ is less than her total traded quantity in $M_2$, i.e., $Q(M_2,b')>Q(M_1,b')$ (since the sum of the total traded quantities of all the bids of $B$ in $M_1$ is equal to the sum of the total traded quantities of all the bids of $B$ in $M_2$ from Lemma \ref{lem:absumQ}).

Now, there can be two cases: (i) $b$ is more competitive than $b'$ or (ii) $b'$ is more competitive than $b$, as per price-time priority. In the first case, since $Q(M_1,b)>Q(M_2,b)$, it follows that $Q(M_2,b) < Q(M_1,b) \leq q_b$. This contradicts the fact that $M_2$ is fair on the bids; this is because a less competitive bid $b'$ is being traded in $M_2$ (since $Q(M_2,b')>Q(M_1,b') \ge 0$ as noted above), while a more competitive bid $b$ is not fully traded. Similarly, in the second case, we show a contradiction to the fact that $M_1$ is fair on the bids. $\hfill\square$


From the above theorem, we have the following corollary.

\begin{theorem} \label{thm:uniquenessTheorem}
        For any two fair and optimal individual-rational-uniform matchings $M_1$ and $M_2$ on the list of bids $B$ and the list of asks $A$, for each order $\omega$, the total traded quantity of $\omega$ in $M_1$ is equal to the total traded quantity of $\omega$ in $M_2$.
\end{theorem}

For each trader, we can compare the total traded quantities of the trader in the matching $M_1$ produced by an exchange with the total traded quantities of the trader in the matching $M_2=\UM(B, A)$ produced by our certified program. If for some trader, the traded quantities do not match, then from Theorem \ref{thm:um} and Theorem \ref{thm:uniquenessTheorem} we know that $M_1$ does not have the desired properties as required by the regulators. On the other hand, if they do match for all traders, then the following theorem states that $M_1$ is fair (Note that uniform and individual-rational properties can be verified directly from the trade prices and clearly the total trade volume of $M_1$ and $M_2$ are the same if the traded quantities are same for each trader).

\begin{theorem} \label{thm:uniquenessTheorem2}
        Given a list of bids $B$ and a list of asks $A$, if $M_1$ is a fair matching and $M_2$ is an arbitrary matching such that for each order $\omega$, the total traded quantity of $\omega$ in $M_1$ is equal to the total traded quantity of $\omega$ in $M_2$, then $M_2$ is fair.
\end{theorem}

The proof follows immediately from  the definition of fairness.

Formalization notes: All the theorems in this section are formalized in the file 'Uniqueness.v' using the above proof ideas.

\section{Demonstration: Automatic Detection of Violations in Real Data}
\label{sec:demonstration}
Please see Appendix~\ref{appendix:demo} for details on our demonstration, where we automate the process of checking violations in trades using verified programs extracted from our formalization. We then use this to find violations in trades of 100 stocks traded on a real exchange on a particular day. Below, we describe our findings.

Out of the 100 stocks we checked, for three stocks our program outputted "Violation detected!". When we closely examined these stocks, we realized that in all of these stocks, a market ask order (with limit price = 0), was not matched by the exchange in its trading output (and these were the only market ask orders in the entire order-book). On the contrary, market bid orders were matched by them. With further investigation, we observed that corresponding to each of these three violations, in the raw data there was an entry of update request in the order-book with a limit price and timestamp identical to the uniform price and the auction time, respectively. It seems highly unlikely that these three update requests were placed by the traders themselves (to match the microsecond time and also the trade price seems very improbable); we suspect this is an exchange's system generated entry in the order-book. We hope that the exchange is aware of this and doing this consciously. When we delete the market asks in the preprocessing stage, no violations are detected. Even if it is not a violation (but a result of the exchange implementing some unnatural rule that we are not aware of), it is fascinating to see that with the help of verified programs we can identify such minute and interesting anomalies which can be helpful for regulating and improving the exchange's matching algorithm.

\section{Related Works and Future Direction}
\label{sec:conclusion}
In an earlier work \cite{icfem}, Sarswat and Singh dealt primarily with single unit trade requests
and thus provided a proof of concept for obtaining verified programs for financial markets.
In the current work, we extend their work to multiple units that results in verified
programs which we run on real market data and establish new uniqueness theorems that enable automatic detection of violation in exchanges as demonstrated in this work.

Passmore and Ignatovich in \cite{PassmoreI17} highlight  the significance, 
opportunities and challenges involved in formalizing financial 
markets. They describe the whole spectrum of financial 
algorithms that need to be verified for ensuring safe and fair markets. 
Iliano {\em et al.} \cite{clf} use concurrent linear logic (CLF) to outline two important properties of a 
continuous trading system. 
There are also some works formalizing 
various concepts from auction theory \cite{welfare,nash,frank}, particularly 
focusing on the Vickrey auction mechanism.

In our opinion, future works should focus on developing a theory for continuous double
auctions for financial markets. Currently the specifications for continuous double auctions
are vague and this is an obstacle for obtaining verified programs.

\section*{Acknowledgment} We wish to thank Mohit Garg for his generous contribution to this work.

\bibliographystyle{plain}
\bibliography{mdsa}

\begin{thebibliography}{10}

\bibitem{git:mdsa}
Coq formalization of mdsa.
\newblock {\verb+https://github.com/suneel-sarswat/dsam+}.

\bibitem{clf}
Iliano Cervesato, Sharjeel Khan, Giselle Reis, and Dragisa Zunic.
\newblock Formalization of automated trading systems in a concurrent linear
  framework.
\newblock In {\em Linearity-TLLA@FLoC}, volume 292 of {\em EPTCS}, pages 1--14,
  2018.

\bibitem{welfare}
Cezary Kaliszyk and Julian Parsert.
\newblock Formal microeconomic foundations and the first welfare theorem.
\newblock In {\em Proceedings of the 7th ACM SIGPLAN International Conference
  on Certified Programs and Proofs}, pages 91--101. ACM, 2018.

\bibitem{nash}
St{\'e}phane Le~Roux.
\newblock Acyclic preferences and existence of sequential nash equilibria: a
  formal and constructive equivalence.
\newblock In {\em International Conference on Theorem Proving in Higher Order
  Logics}, pages 293--309. Springer, 2009.

\bibitem{mcafee1992}
R~Preston McAfee.
\newblock A dominant strategy double auction.
\newblock {\em Journal of economic Theory}, 56(2):434--450, 1992.

\bibitem{NiuP13}
Jinzhong Niu and Simon Parsons.
\newblock Maximizing matching in double-sided auctions.
\newblock In {\em International conference on Autonomous Agents and Multi-Agent
  Systems, AAMAS '13, Saint Paul, MN, USA, May 6-10, 2013}, pages 1283--1284,
  2013.

\bibitem{PassmoreI17}
Grant~Olney Passmore and Denis Ignatovich.
\newblock Formal verification of financial algorithms.
\newblock In {\em 26th International Conference on Automated Deduction,
  Proceedings}, volume 10395 of {\em Lecture Notes in Computer Science}, pages
  26--41. Springer, 2017.

\bibitem{icfem}
Suneel Sarswat and Abhishek~Kr Singh.
\newblock Formally verified trades in financial markets.
\newblock In Shang{-}Wei Lin, Zhe Hou, and Brendan Mahoney, editors, {\em
  Formal Methods and Software Engineering - 22nd International Conference on
  Formal Engineering Methods, {ICFEM} 2020, Singapore, Singapore, March 1-3,
  2021, Proceedings}, volume 12531 of {\em Lecture Notes in Computer Science},
  pages 217--232. Springer, 2020.

\bibitem{equations}
Matthieu Sozeau and Cyprien Mangin.
\newblock Equations reloaded: High-level dependently-typed functional
  programming and proving in coq.
\newblock {\em Proceedings of the ACM on Programming Languages}, 3(ICFP):1--29,
  2019.

\bibitem{frank}
Emmanuel~M. Tadjouddine, Frank Guerin, and Wamberto~Weber Vasconcelos.
\newblock Abstracting and verifying strategy-proofness for auction mechanisms.
\newblock In {\em DALT}, volume 5397 of {\em Lecture Notes in Computer
  Science}, pages 197--214. Springer, 2008.

\bibitem{WurmanWW98}
Peter~R. Wurman, William~E. Walsh, and Michael~P. Wellman.
\newblock Flexible double auctions for electronic commerce: theory and
  implementation.
\newblock {\em Decision Support Systems}, 24(1):17--27, 1998.

\bibitem{ZhaoZKP10}
Dengji Zhao, Dongmo Zhang, Md~Khan, and Laurent Perrussel.
\newblock Maximal matching for double auction.
\newblock In {\em Australasian Conference on Artificial Intelligence}, volume
  6464 of {\em Lecture Notes in Computer Science}, pages 516--525. Springer,
  2010.

\end{thebibliography}

\newpage

\appendix\label{appendix}
\section{Demonstration on real data.}
\label{appendix:demo}

In this section, we demonstrate the practical applicability of our work. For this, we procured real data from a prominent stock exchange. This data consists of order-book and trade-book of everyday trading for a certain number of days. For our demonstration, we considered trades for the top 100 stocks (as per their market capitalizations) of a particular day. For privacy reasons, we conceal the real identity of the traders, stocks and the exchange by masking the stock names (to s1 to s100) and the traders' identities. We also converted the timestamps appropriately into natural numbers (which keeps the time in microseconds, as in the original data). Furthermore, the original data has multiple requests with the same order id; this is because some traders update or delete an existing order placed by them before the double sided auction is conducted. In our preprocessing, we just keep the final lists of bids and asks in the order-book that participate in the auction. Furthermore, there are certain market orders, i.e., orders that are ready to be traded at any available price, which effectively means a limit price of zero for an ask and a limit price of infinity for a bid; in the preprocessing we set these limit prices to zero and the largest OCaml integer, respectively.

We then extracted the verified OCaml programs and ran them on the processed market data. The output trades of  the verified code were then compared with the actual trades in the trade-book from the exchange. From the uniqueness theorems in the Section~\ref{sec:uniqueness}, we know that if the total trade quantity of each order in these two matchings are equal, then the matching produced by the exchange has the desired properties (if it is uniform and IR which can be checked trivially by looking at the prices in the trade-book). We also know that if they are not equal for some trader, then the matching algorithm of the exchange does not have the requisite desired properties (or there is some error in storing or reporting the order-book or the trade-book accurately).

The processed data and the relevant programs for this demonstration are available at ~\cite{git:mdsa}. The extracted OCaml programs of the functions required for this demonstration are stored in a separate file named `certified.ml'. The input bids, asks and trades of each stock are in `s.bid', `s.ask' and `s.trade' files, where `s' is the masked id for that stock. For example, file `s1.bid' contains all the bids for the stock `s1'. To feed the inputs to the verified program and to print the output of the certified program, we have written two OCaml scripts: create.ml and compare.ml. The create.ml script feeds inputs (lists of bids and asks) to the UM process, and then prints its output matching $M$. The compare.ml script compares the matching produced by the UM process $M$ with the actual trades $M_\text{EX}$ in the exchange trade-book. If the total trade quantity for all the traders in $M$ matches with that of the total trade quantity in $M_\text{EX}$, then the compare.ml script outputs "Matching does not violate the guidelines". If for some bid (or ask) the total trade quantity of $M$ and $M_\text{EX}$ does not match, then the program outputs "Violation detected!".

\section{Proof details for Theorem~\ref{thm:FOB}.}
\label{fairappendix}

For a given matching $M$ on the list of bids $B$ and the list of asks $A$, we prove the main theorem by broadly establishing the following properties.
\begin{enumerate}
\item $\text{For all } m \in f(M,B,0),\ p(a_m) \le p(b_m)$.
\item $B_{f(M,B,t)} \subseteq B$.
\item $A_{f(M,B,t)} \subseteq A_M$.
\item $\text{For all } b \in B_{f(M,B,t)},\ Q(b,f(M,B,t)) \le q_b$.
\item $\text{For all } a \in A_{f(M,B,t)},\ Q(a,f(M,B,t)) \le q_a$.
\item $\text{For all } a \in A_M,\ Q(a,f(M,B,0)) = Q(a,M)$.
\item $\text{For all } a\in A_M, \text{ if } Q(B) \ge Q(M)+t \text{, then } Q(a,M) = Q(a, f(M, B, t))$.
\item $\text{For all } b_1, b_2 \in B_{f(M,B,0)}, \text{ if }b_1 \text{ is more competitive than }b_2 \text{, then } Q(b_1, f(M,B,0)) = q(b_1)$.
\item $\text{For all } b_1, b_2 \in B_{f(M,B,t)}, \text{ if }b_1 \text{ is more competitive than }b_2  \text{, then } Q(M) > (q_{b_1} - t)$.
\item $\text{If }Q(M) \ge (q_b - t) \text{, then } Q(b, f(M,b::B,t)) = (q_b - t)$.
\end{enumerate} 

Observe that, the $\FOB$ procedure calls subroutine $f$ with $t=0$. However, if we try to prove the properties of $\FOB$ using induction while fixing $t=0$, we get a weaker induction hypothesis. This is because in one of the recursive cases, the value of $t$ is not zero. On the other hand, some results holds true only for $t=0$. Therefore, it is helpful to separately prove the results essential for the main theorem and identify how the correctness of each of these results is sensitive to the different values of $t$

In the above list of properties, (1) - (5) are needed to prove that $M_\beta$ is matching and are closely linked to the five properties of Definition~\ref{def:matching}. Observe that while the properties (2) - (5) are true for arbitrary values of $t$, property (1) is special case with term $t=0$. The property (1) above says that all the bid-ask pairs produced by $f$ are matchable only when the function $f$ is called with $t=0$. For see this, consider the function call $f(m::M',b::B',q_b)$ with $t = q_b$. Also assume that none of the bids of $B'$ is matchable with the ask of $m$(i.e., $\forall b'\in B', p(b')< p(a_m)$). In this case the most competitive bid $b$ is already fully traded, as indicated by $t$, and hence the function $f$ will try to match $a_m$ with some other $b'\in B'$, which will not be matchable.

By substituting $t=0$ in (7), we can prove property (6) which states that function $\FOB$ does not change the filled quantity for any ask $a$ appearing in the original matching $M$. Similarly, by using properties (9) and (10) we can prove property (8) above, which states that a more competitive bid $b_1$ is fully traded before starting any transaction involving a less competitive bid $b_2$. Also note that similar to earlier, property (8) is only true for $t=0$, the properties (9) and (10) are statements about the properties of $f(M,B,t)$ for an arbitrary $t$.

\section{A Maximum Matching Mechanism}\label{sec:maxappendix}
In  Section \ref{sec:uniform}, we noted that to obtain a maximum matching on certain inputs we may have to compromise the uniform properties. We now describe a matching mechanism which produces a maximum trade volume matching. We can then use the function $\FOA$ to make it fair.

\begin{definition} \textcolor{gray}{Maximum Matching ($\MM$)}.
\begin{align*}
&\MM( B, A) = \FOA(\stackrel{\mathbb{A}}{\uparrow} f_m(\stackrel{\mathbb{B}}{\uparrow} B, \stackrel{\mathbb{A}}{\downarrow} A,0,0) , \stackrel{\mathbb{A}}{\uparrow} A) \\
&\text{where }  f_m(B,A,t_b,t_a) =\\
         &\begin{cases}
         nil  &\text{if $B = nil$ or $A = nil$} \\
             f(b::B',A',t_b,0) & \text{if $p_b < p_a$} \\
                (b,a,q_a - t_a,p_a)::f_m(B',A',0,0) & \text{if $q_a - t_a=q_b - t_b \text{ and } p_b \ge p_a$} \\
                (b,a,q_a - t_a,p_a)::f_m(b::B',A',q_b + q_a - t_a,0) & \text{if $q_a - t_a < q_b - t_b \text{ and } p_b \ge p_a $} \\
                (b,a,q_b - t_b,p_a)::f_m(B',a::A',0,q_a + q_b - t_a) & \text{if $q_a - t_a > q_b - t_b \text{ and } p_b \ge p_a$}
        \end{cases}
\end{align*}
where $B=b::B'$ when $B \neq nil$ and $A = a::A'$ when $A \neq nil$.
\end{definition}

\emph{Description}. Procedure $\MM(B,A)$ first sorts the list of bids $B$ in decreasing order of their competitiveness and the list of asks $A$ in increasing order of their competitiveness and calls the $f_m$ procedure on these sorted lists. Similar to $\UM$ procedure, we use parameters $t_b$ and $t_a$ that inform the recursive procedure $f_m$ that the top bid $b$ and the top ask $a$ have effective quantities $q_b - t_b$ and $q_a - t_a$, respectively. In each iteration, the function $f_m$ first check if the most competitive bid $b$ is matchable to the least competitive ask $a$; If they are matchable (i.e $p_b \ge p_a$), then a transaction between $b$ and $a$ with quantity $q = \min\{q_b - t_b,q_a - t_a\}$ and price $p_a$ is created and the function recursively proceeds on smaller inputs by appropriately updating the filled trade quantities of the top bid and ask; If $b$ and $a$ are not matchable (i.e. $p_b<p_a$) the function ignores the current ask $a$ and proceeds recursively on smaller input. If either of the lists $B$ or $A$ becomes nil, the function $f_m$ terminates and outputs a matching between the lists of bids $B$ and asks $A$.

\begin{description}
\item \emph{Remark}. The function $f_m$ always tries to match the unfilled trade quantities of the most competitive bid $b$ in $B$ against the unfilled quantities of the first matchable least competitive ask in $A$. It is important to note that due to this difference in operation, now we cannot claim that in each successive recursive call of $f_m$, when it outputs a transaction, the trade price assigned by it (i.e. $p_a$) is greater than or equal to the limit prices of all the asks paired so far (due to the reversed order or sorting of $A$). Hence, unlike $\UM$ process we are not sure about the possibility of finding a common price that can be used to make it uniform as well as individual-rational simultaneously. Moreover, the procedure $f_m$ outputs a fair on bids matching, it might not be fair on asks. Therefore, the procedure $\MM$ calls the procedure $\FOA$ on the output of $f_m$ and the lists of asks $A$ (now, sorted in decreasing order of their competitiveness). As a result, $\MM$ outputs a fair matching on the lists of bids $B$ and the asks $A$.
\end{description}

We now discuss some important results leading to the main theorem of this section which states that for an arbitrary list of bids $B$ and list of asks $A$, where $B$ is sorted by decreasing order of the competitiveness while $A$ is sorted in increasing order of the competitiveness, the function $f_m$ produces a matching which is maximum by trade volume among all the matchings on $B$ and $A$.

\begin{lemma} \label{lem:max_MM'}
If $M$ is a matching on the list of bids $B=\stackrel{\mathbb{B}}{\uparrow}(b::B')$ and  the list of asks $A= \stackrel{\mathbb{A}}{\downarrow} (a::A')$, and $Q(M) \ge q$, where $q= \min \{q_b, q_a\}$, then there exists a matching $M'$ on $B$ and $A$, with $Q(M') = Q(M)$ and $Q(b,M') \ge q$.
\end{lemma}
\emph{Proof Idea}. For the given $M$, $B$ and $A$ consider $M' = \FOB(\stackrel{\mathbb{B}}{\uparrow} M, \stackrel{\mathbb{B}}{\uparrow} b::B')$. Since $\FOB$  does not change the overall trade volume we have $Q(M') \ge q$. Now since $M'$ is fair on bids  with total trade volume higher than the available quota $q$ of both $a$ and $b$, it must fill at least $q$ quantity of $b$ before trading the other less competitive bids from $B'$. Hence, we have  $Q(b,M') \ge q$.  \hfill $\square$

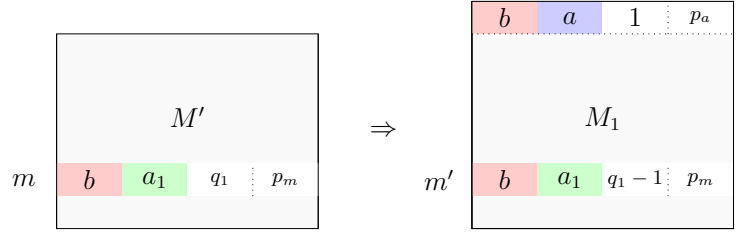
\begin{figure}[ht]
\begin{center}

\resizebox {.7\textwidth} {!} {

\begin{tikzpicture}[font=\Large]

 \draw [fill=gray!5] (0,0) -- (4,0) -- (4,3) -- (0,3)--cycle;
 \fill [fill=red!20] (0,.5) -- (1,.5) -- (1,1) -- (0,1)--cycle;
 \fill [fill=green!20] (1,.5) -- (2,.5) -- (2,1) -- (1,1)--cycle;
 \draw (2, 2) node[below] {$M'$};
 \draw (1.5, .75) node {$a_1$};
 \draw (.5, .75) node {$b$};

 \fill [fill=white] (2,.5) -- (3,.5) -- (3,1) -- (2,1)--cycle;
 \draw (2.5, .75) node {\small{$q_1$}};

 \fill [fill=white] (3,.5) -- (4,.5) -- (4,1) -- (3,1)--cycle;
 \draw (3.5, .75) node {\small{$p_m$}};
 \draw [dotted] (3,.5) -- (3,1);

 \draw (4,0) -- (4,3) --cycle;
 \draw (0,0) -- (0,3) --cycle;

 \draw (-.5, .75) node {$m$};

 \draw [fill=gray!5,xshift=2.5in] (0,0) -- (4,0) -- (4,3.5) -- (0,3.5)--cycle;
 \fill [fill=red!20,xshift=2.5in] (0,.5) -- (1,.5) -- (1,1) -- (0,1)--cycle;
 \fill [fill=green!20,xshift=2.5in] (1,.5) -- (2,.5) -- (2,1) -- (1,1)--cycle;
 \draw (2, 2) node[below,xshift=2.5in] {$M_1$};
 \draw (1.5, .75) node[xshift=2.5in] {$a_1$};
 \draw (.5, .75) node[xshift=2.5in] {$b$};

 \fill [fill=white,xshift=2.5in] (2,.5) -- (3,.5) -- (3,1) -- (2,1)--cycle;
 \draw (2.5, .75) node[xshift=2.5in] {\small{$q_1 - 1$}};

 \fill [fill=white,xshift=2.5in] (3,.5) -- (4,.5) -- (4,1) -- (3,1)--cycle;
 \draw (3.5, .75) node[xshift=2.5in] {\small{$p_m$}};
 \draw [dotted,xshift=2.5in] (3,.5) -- (3,1);

 \draw [xshift=2.5in] (4,0) -- (4,3.5) --cycle;
 \draw [xshift=2.5in] (0,0) -- (0,3.5) --cycle;

 \fill [fill=red!20,xshift=2.5in] (0,.5+2.5) -- (1,.5+2.5) -- (1,1+2.5) -- (0,1+2.5)--cycle;
 \fill [fill=blue!20,xshift=2.5in] (1,.5+2.5) -- (2,.5+2.5) -- (2,1+2.5) -- (1,1+2.5)--cycle;
 \fill [fill=white,xshift=2.5in] (2,.5+2.5) -- (3,.5+2.5) -- (3,1+2.5) -- (2,1+2.5)--cycle;
 \fill [fill=white,xshift=2.5in] (3,.5+2.5) -- (4,.5+2.5) -- (4,1+2.5) -- (3,1+2.5)--cycle;
 \draw [xshift=2.5in] (0,0) -- (4,0) -- (4,3.5) -- (0,3.5)--cycle;
 \draw [dotted,xshift=2.5in] (0,3) -- (4,3);

 \draw [xshift=2.5in] (.5, 3.25) node {$b$};
 \draw [xshift=2.5in] (1.5, 3.25) node {$a$};
 \draw [xshift=2.5in] (2.5, 3.25) node {$1$};
 \draw [xshift=2.5in] (3.5, 3.25) node {\small{$p_a$}};

 \draw [dotted,xshift=2.5in] (3,3) -- (3,3.5);

 \draw [xshift=2.5in] (-.5, .75) node {$m'$};
 \draw (5, 1.5) node {{$\Rightarrow$}};

\end{tikzpicture}
}
\end{center}
\caption{In the above figure the matching $M_1$ is obtained from the matching $M'$. Total trade volume in both $M'$ and $M_1$ is equal. Furthermore, the trade quantity between $a$ and $b$ in $M_1$ is one more than that in $M'$. }
\label{fig:surgery2}

\end{figure}

\begin{lemma} \label{lem:max_M'M''}
If $M'$ is a matching on the list of bids $B= \stackrel{\mathbb{B}}{\uparrow} (b::B')$ and  the list of asks $A= \stackrel{\mathbb{A}}{\downarrow} (a::A')$, with $p_b \ge p_a$ and $Q(b,M') \ge q$, where $q= \min \{q_b, q_a\}$, then there exists a matching $M''$ on $B$ and $A$, with $Q(M'') = Q(M')$ and $Q(a \leftrightarrow b, M'') = q$.
\end{lemma}
\emph{Proof Idea}. Note that while the term $Q(a \leftrightarrow b, M'')$ represents the total volume of trade between  $b$ and $a$ in $M''$, the quantity $q$ represents the maximum possible trade between $b$ and $a$ in any arbitrary matching $M$ provided $p_b \ge p_a$. Now assume that we have $Q(a \leftrightarrow b, M') = q-k$, where $k$ represents the trade deficit between $b$ and $a$ particular to $M'$. The existence of a matching $M''$ with no trade deficit can be shown using mathematical induction on the value of $k$. Again, at the core of this inductive argument is a method by which we generate another matching $M_1$ from $M'$ with same trade volume while reducing the trade deficit among $a$ and $b$ by one unit. We now consider the following two possibilities (1) $Q(a,M') < q$, and (2) $Q(a, M') \ge q$. In the case (1) if $k>0$ we can prove that $\exists m \in M', m = (b, a_1, q_1, p_m)$, where $a_1 \neq a$. Now consider $M_1 = (b,a,1,p_a)::M'[m \mapsto m']$, where $m' = (b, a_1, q_1 - 1,p_m)$ (see Figure \ref{fig:surgery2}). It is easy to see that $M_1$ is a matching with  $Q(a \leftrightarrow b, M_1) = q-k-1$ while $Q(M_1) = Q(M')$. In the case (2), where  $Q(a, M') \ge q$, if $k >0$ then it can be shown that $\exists \; m_1 \; m_2 \in M', m_1= (b, a_1, q_1, p_1) \text{ and } m_2 = (b_1, a, q_2, p_2)$, where $b \neq b_1$ and $ a \neq a_1$. Moreover, since $\stackrel{\mathbb{B}}{\uparrow} b::B'$, $\stackrel{\mathbb{A}}{\downarrow} a::A'$, and $m_1$ and $m_2$ are valid transactions we can conclude that $p_b \ge p(b_1) \ge p_a \ge p(a_1)$.  Furthermore, we define $M_1 = (b,a,1,p_a)::(b_1,a_1,1,p(a_1))::M'[m_1 \mapsto m_1', m_2 \mapsto m_2']$, where $m_1'= (b, a_1, q_1 - 1, p_1) $ and $m_2'=  (b_1, a, q_2 - 1, p_2)$. Note that $Q(a \leftrightarrow b, M_1) = q-k-1$ while $Q(M_1) = Q(M')$. This process can be repeated to get a $M''$ where the trade deficit is zero (i.e.  $Q(a \leftrightarrow b, M'') = q$).  \hfill $\square$

\begin{theorem} \label{thm:MM}
Let $M$ be an arbitrary matching on the given list of bids $B$ and list of asks $A$, and $M_m = f_m( \stackrel{\mathbb{B}}{\uparrow B}, \stackrel{\mathbb{A}}{\downarrow} A, 0, 0)$ then $M_m$ is a matching on $B$ and $A$ with $Q(M_m) \ge Q(M)$.
\end{theorem}
\emph{Proof Outline}. The proof of this theorem is using induction on the sum $(|B|+|A|)$, however due to the reasons similar in the proof of Theorem \ref{thm:UMopt} we prove this theorem by proving the following claim which incorporates the changing values of $t_a$ and $t_b$.
\begin{itemize}
\item  $\forall M,  \text{Matching } M \; ((id_b,\tau_b,q_b - t_b,p_b)::B) \; ((id_a,\tau_a, q_a - t_a, p_a)::A)\rightarrow 
                  Q(f_m(\stackrel{\mathbb{B}}{\uparrow} b::B,\stackrel{\mathbb{A}}{\downarrow} a::A,t_b,t_a)) \ge Q(M)$
\end{itemize}
We prove this result using induction on the value $(|b::B| + |a::A|)$, and unfolding the definition of $f_m$ in each branch of the computation. We get the following induction hypothesis (IH),  where $B'_{-t_b}$ and $A'_{-t_a}$ are the list of bids and list of asks obtained after updating the unfilled quotas of the top bid and top ask as per the values of $t_b$ and $t_a$ respectively.

\begin{itemize}
\item  IH: $\forall \;B'\; A'\; t_a\; t_b \; M_0,\, |B'|+|A'| < |b::B|+|a::A| \rightarrow  (\text{Matching}\; M_0 \; B'_{-t_b}  A'_{-t_a} \\  \rightarrow Q(f_m(\stackrel{\mathbb{B}}{\uparrow B'},\stackrel{\mathbb{A}}{\downarrow A'}, t_b,t_a)) \ge Q(M_0)) $
\end{itemize}

We first consider the case when $p_b < p_a$. In this case $f_m(b::B,a::A,t_b,t_a)$ reduces to $f_m(b::B,A,t_b,0)$ and hence produces a matching on $b::B$ and $A$. Moreover, since $p_a > p_b$ the ask $a$ cannot participate in any matching between $b::B$ and $a::A$. Hence we can conclude that $M$ is also a matching between $b::B$ and $A$. Therefore using induction hypothesis we have $ Q(f_m(\stackrel{\mathbb{B}}{\uparrow} b::B,\stackrel{\mathbb{A}}{\downarrow} a::A,t_b,t_a)) = Q(f_m(\stackrel{\mathbb{B}}{\uparrow} b::B,\stackrel{\mathbb{A}}{\downarrow} A,t_b,0)) \ge Q(M)$, which proves the result in this case. Now consider the case when $a$ and $b$ are matchable (i.e. $p_b \ge p_a$). Let $q = \min \{q_b - t_b, q_a - t_a\}$, and $Q(M) < q$. In this case, the theorem statement is true since our algorithm produces a matching of total trade volume greater than or equal to $q$. Now consider the second case when $Q(M) \ge q$. In this case, using Lemma \ref{lem:max_MM'} and Lemma \ref{lem:max_M'M''}, we prove that there exists a matching $M''$ between the list of bids  $(id_b,\tau_b, q_b - t_b, p_b)::B$ and the list of asks $(id_a,\tau_a, q_a - t_a, p_a)::A$, such that $Q(M'') = Q(M)$ and $Q(a \leftrightarrow b, M'') = q$.  Note that we now have a matching $M''$ with the same total trade volume as $M$ but having maximum possible trade between the top bid $b$ and top ask $a$. Therefore, we complete the proof of this theorem by proving $Q(f_m(\stackrel{\mathbb{B}}{\uparrow} b::B,\stackrel{\mathbb{A}}{\downarrow} a::A,t_b,t_a)) \ge Q(M'')$.  We now prove this result in each of the following three cases.

\noindent $\triangleright$ \emph{C 1}. In this case, when $q_a - t_a= q_b - t_b$, we have $ Q(f_m(\stackrel{\mathbb{B}}{\uparrow} b::B,\stackrel{\mathbb{A}}{\downarrow} a::A,t_b,t_a)) =  q + Q(f_m(B,A,0,0)) \ge q + Q(M_0) = Q (M'')$, where $M_0$ is a  matching between $B$ and $A$.  Note that we can claim $q + Q(f_m(B,A,0,0)) \ge q + Q(M_0)$ since $Q(f_m(B,A,0,0)) \ge Q(M_0)$ from the induction hypothesis.

\noindent $\triangleright$ \emph{C 2}. In this case $q_a - t_a > q_b - t_b$, and we have  $ Q(f_m\stackrel{\mathbb{B}}{(\uparrow} b::B,\stackrel{\mathbb{A}}{\downarrow} a::A,t_b,t_a)) =  q + Q(f_m(B,a::A,0,t_a + q_b - t_b)) \ge q + Q(M_0) = Q (M'')$, where  $M_0$ is a matching between $B$ and $(id_a,\tau_a,(q_a - t_a - (q_b - t_b)),p_a)::A$. Note that in this case  $q=(q_b - t_b)$ and again we can claim $ q + Q(f_m(B,a::A,0,t_a + q_b - t_b)) \ge q + Q(M_0)$ because we know $ Q(f_m(B,a::A,0,t_a + q_b - t_b)) \ge Q(M_0)$ from the induction hypothesis.

\noindent $\triangleright$ \emph{C 3}. In this case $q_a - t_a < q_b - t_b$ and we have  $ Q(f_m(\stackrel{\mathbb{B}}{\uparrow} b::B,\stackrel{\mathbb{A}}{\downarrow} a::A,t_b,t_a)) =  q + Q(f_m(b::B,A,t_b + q_a - t_a,0))  \ge q + Q(M_0) = Q (M'')$, where  $M_0$ is a matching between   $(id_b,\tau_b,(q_b - t_b - (q_a - t_a)), p_b)::B$ and $A$. Again in this case $q + Q(f_m(b::B,A,t_b + q_a - t_a,0))  \ge q + Q(M_0)$ follows from $Q(f_m(b::B,A,t_b + q_a - t_a,0))  \ge Q(M_0)$  from the induction hypothesis.   \hfill $\square$

\begin{theorem} \label{thm:mmapendix}
For a given list of bids $B$ and the list of asks $A$, $M = \MM (B ,A)$ is a fair, individual-rational and maximum volume matching on $B$ and $A$.
\end{theorem}

\end{document}